\documentclass[10pt,twocolumn]{article}
\usepackage{times}

\usepackage{cite}
\usepackage{url}
\usepackage{multirow}
\usepackage{amsmath}
\usepackage{graphicx}
\usepackage{subfigure}
\usepackage{algorithm}
\usepackage{algorithmic}
\usepackage{tabularx}
\usepackage{color}

\usepackage{ulem}
\usepackage{textcomp}

\newcommand{\tabincell}[2]{\begin{tabular}{@{}#1@{}}#2\end{tabular}}

\begin{document}

\title{Write-Optimized and Consistent RDMA-based NVM Systems}

\author{
{  Xinxin Liu,
    Yu Hua,
    Xuan Li,
    Qifan Liu
}\\
Huazhong University of Science and Technology \\
}

\date{}
\maketitle

\begin{abstract}
In order to deliver high performance in cloud computing, we generally exploit and leverage RDMA (Remote Direct Memory Access) in networking and NVM (Non-Volatile Memory) in end systems. Due to no involvement of CPU, one-sided RDMA becomes efficient to access the remote memory, and NVM technologies have the strengths of non-volatility, byte-addressability and DRAM-like latency. In order to achieve end-to-end high performance, many efforts aim to synergize one-sided RDMA and NVM. Due to the need to guarantee Remote Data Atomicity (RDA), we have to consume extra network round-trips, remote CPU participation and double NVM writes. In order to address these problems, we propose a zero-copy log-structured memory design for Efficient Remote Data Atomicity, called Erda. In Erda, clients directly transfer data to the destination address at servers via one-sided RDMA writes without redundant copy and remote CPU consumption. To detect the incompleteness of fetched data, we verify a checksum without client-server coordination. We further ensure metadata consistency by leveraging an $8$-byte atomic update in the hash table, which also contains the address information for the stale data. When a failure occurs, the server properly restores to a consistent version. Experimental results show that compared with Redo Logging (a CPU involvement scheme) and Read After Write (a network dominant scheme), Erda reduces NVM writes approximately by $50\%$, as well as significantly improves throughput and decreases latency.
\end{abstract}

\section{Introduction}\label{sec:introduction}

Cloud computing requires high performance in both network transmission and local I/O throughput. Remote direct memory access (RDMA) technologies have become more important for cloud computing~\cite{shan2017distributed,guo2016rdma}. RDMA allows to directly access remote memory via bypassing kernel and zero memory copy, thus providing high bandwidth and low latency for remote memory accesses~\cite{tsai2017lite}. Moreover, non-volatile memory (NVM) technologies have the strengths of non-volatility, byte-addressability, high density and DRAM-class latency in end systems. NVM can be directly accessed through the network and local memory bus with RDMA protocol and CPU load/store instructions~\cite{zuo2018write}. Many schemes thus synergize RDMA and NVM to deliver end-to-end high performance~\cite{hu2018persistence,kim2018hyperloop,lu2017octopus,islam2016high,shan2017distributed,zhang2015mojim}.

Since one-sided RDMA operations (read, write and atomic) do not involve remote CPU but two-sided (send and recv) do, one-sided primitives provide higher bandwidth and lower latency than two-sided one. For CPU-intensive workloads, even if one-sided primitives require more network round-trips than two-sided primitives, one-sided primitives are still faster than two-sided primitives~\cite{wei2018deconstructing,lu2017octopus}. However, using one-sided RDMA to access remote NVM becomes inefficient due to the challenges of guaranteeing \textbf{Remote Data Atomicity (RDA)}: Incomplete writes from failures are durable in NVM, resulting in inconsistent data. The server is unaware of the incomplete and invalid data in NVM due to no CPU involvement in the context of the one-sided RDMA operations. The client is also unaware of the possible data loss in the server, because the returned ACK of RDMA write from the server merely means that the data have reached the volatile cache of the server NIC, and possibly fail to be flushed into NVM.

However, many existing RDMA-based NVM systems overlook RDA and become inefficient in system performance~\cite{lu2017octopus,hu2018persistence,islam2016high,chen2019scalable}. For example, in collect-dispatch transaction in Octopus~\cite{lu2017octopus}, a coordinator uses one-sided RDMA write to update the write sets in participants, which are unaware of the incomplete data without the CPU involvements of participants. Hence, if a failure occurs before the written data are fully flushed from the volatile cache of the participants NIC into NVM, the write sets will be partially applied and durable in NVM, which is neither the ``old'' nor the ``new'' version, thus becoming inconsistent~\cite{zuo2018write,shan2017distributed}.

In order to guarantee RDA, some schemes leverage an extra RDMA read operation after RDMA write to force data to be persistent and integrated~\cite{ChetDouglas_RDMA_with_PM,ChetDouglas_Intel_Perspective,kim2018hyperloop}. Undo logging, redo logging and copy-on-write (COW) are consistency mechanisms and have been widely used in persistent memory systems~\cite{shan2017distributed,nguyen2018picl,ogleari2018steal,nam2019write,dulloor2014system}. There also exist some RDMA-based NVM systems that ensure RDA by CPU involvement~\cite{yang2019orion,shan2017distributed,zhang2015mojim}. However, these solutions unfortunately fail to be efficient due to the following problems.

\textbf{High Network Overheads.} The schemes that leverage an extra RDMA read operation after RDMA write cause extra network round-trips for each RDMA write, resulting in high network overheads.

\textbf{High CPU Consumption.} Undo/redo logging and COW require the remote CPU to control operation sequence. However, CPU involvement decreases the benefits of using one-sided RDMA operations that don't need the consumption of remote CPU when accessing the remote memory.

\textbf{Double NVM Writes.} Some CPU involvement solutions need to first check the written data in persistent log regions or buffers, and then apply them into the destination storage. These operations essentially require double NVM writes, consuming the limited NVM endurance. More NVM writes also cause higher latency than reads.

In order to address these problems, we propose Erda (Efficient Remote Data Atomicity) that is a zero-copy log-structured memory design. Erda guarantees RDA for one-sided RDMA writes to NVM without extra network round-trips, remote CPU consumption and double NVM writes. In Erda, an object with a CRC checksum inside is the basic unit of access operations. For the update operation from clients to servers, the metadata in a hash table are modified with an $8$-byte atomic write, and then the object is directly transferred from clients to the destination storage at servers without redundant buffer and server CPUs, thus reducing the amount of write operations approximately by $50\%$, compared with undo/redo logging and COW. The incompleteness of the written object will be detected by subsequent read requests via verifying checksums. Once the verification results show that the fetched object is incomplete, clients will re-read the previous version of the object, whose address information is also contained in the hash table, to ensure the consistency and atomicity of the fetched object. At the same time, servers are notified about the inconsistency and properly restore to a consistent version. Specifically, we have the following contributions:

\textbf{RDA Solution with Low Overheads.} We investigate the Remote Data Atomicity (RDA), and find that exiting RDMA-based NVM systems either overlook RDA or guarantee RDA at high overheads. Exiting solutions need high network overheads, high CPU consumption and double NVM Writes. Our proposed scheme is able to guarantee RDA and improve exiting RDA solutions with low overheads.

\textbf{Cost-efficient Synergized Design.} We propose a zero-copy log-structured memory design, named Erda, which guarantees RDA without extra network round-trips, remote CPU consumption and redundant copy. In Erda, we allow clients to directly write objects to the destination address at servers without buffer and copy. Subsequent read requests will detect the incompleteness of fetched objects without client-server coordination.

\textbf{Evaluation and Open-Source Codes.} We conduct experimental evaluation to exhibit the efficiency of Erda. Evaluation results demonstrate that compared with Read After Write and Redo Logging schemes, Erda significantly improves throughput and decreases latency, as well as reduces NVM writes approximately by $50\%$. The source codes are released for public use at \emph{https://github.com/csXinxinLiu/Erda}.

The rest of this paper is organized as follows.  We present the background of RDMA networking and non-volatile memory in Section~\ref{sec:background}. Section~\ref{sec:design} shows our design. Section~\ref{sec:implementation} shows the implementation details of Erda. The experimental results are shown in Section \ref {sec:evaluation}. We present the related work in Section~\ref{sec:related work}. Finally, we conclude this paper in Section~\ref{sec:conclusion}.

\section{Background}\label{sec:background}
In this Section, we present the background of remote direct memory access (RDMA) and non-volatile memory (NVM).

\subsection{RDMA Networking}
Remote direct memory access (RDMA) bypasses kernel and supports zero memory copy, thus providing extremely high bandwidth and low latency for remote memory accesses~\cite{tsai2017lite}. RDMA has two kinds of primitives, i.e., one-sided and two-sided. One-sided primitives include RDMA read, RDMA write and atomic operations, which do not involve remote CPU when accessing the remote memory. Furthermore, two-sided primitives, such as RDMA send and RDMA recv, are similar to socket programming. Two-sided operations are served by the remote CPU, which must poll RDMA messages and process them.

RDMA is well-known for one-sided primitives, which provide higher bandwidth and lower latency than two-sided RDMA, especially when the remote server is busy~\cite{wei2018deconstructing,lu2017octopus}. Furthermore, there is a trend that one-sided primitives are becoming more and more fast and scalable for recent generations of RNICs, e.g., ConnectX-4 and ConnectX-5~\cite{wei2018deconstructing}.

\subsection{Non-Volatile Memory}

Non-volatile memory (NVM) technologies, such as $3$DXPoint~\cite{3dxpoint} and PCM~\cite{wong2010phase}, have the strengths of non-volatility, byte-addressability, high density and DRAM-class latency. Hence, NVMs are promising candidates of next-generation main memory and caches~\cite{yang2013memristive}, as well as complements to current external storages such as flash-based SSDs~\cite{liu2017durable}.

However, NVMs with different materials have some common limitations. First, NVMs have asymmetric properties of writes and reads. For example,  NVM writes consume higher energy than reads, and also cause higher latency ($3$ -- $8X$) than reads~\cite{yue2013accelerating,yang2013memristive,zuo2018writetpds}. Second, NVMs generally suffer from limited write endurance~\cite{zhou2009durable,yang2013memristive,qureshi2009enhancing}. Therefore, it is necessary to reduce the amount of write operations in NVM systems.

NVMs have the non-volatile capability. They keep the contents across crash or power failure. Therefore, NVM systems must consider consistency mechanism to avoid data corruption. It is well-recognized that the failure atomicity unit for NVM is $8$ bytes, because byte-addressable NVMs are accessible through the memory bus~\cite{lee2017wort,oukid2016fptree,yang2015nv}. If the size of the updated data is larger than the $8$-byte failure-atomic write granularity, existing mechanisms, such as undo/redo logging and copy-on-write (COW), are employed to maintain consistency~\cite{shan2017distributed,nguyen2018picl,ogleari2018steal,nam2019write,dulloor2014system}. Undo logging needs to append old data into an undo log first, and then updates in-place. Redo logging first appends new data in a redo log, and then updates the old data. COW creates a copy and then performs updates on the copy.
\begin{figure}[t]
  \centering
    \includegraphics [width=0.48\textwidth]{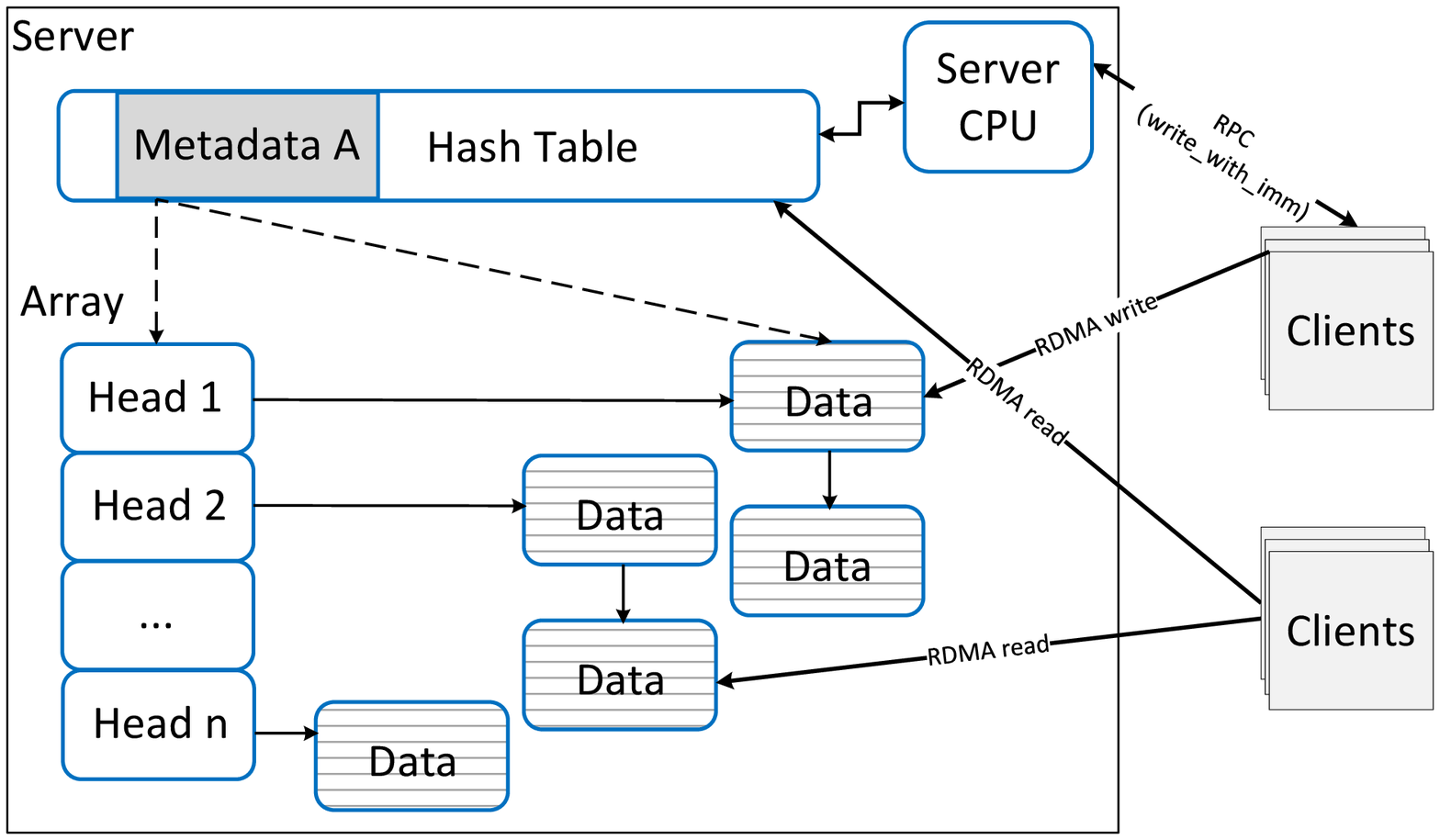}
\vspace{-5px}
    \caption{\label{fig:architecture} The overall architecture of Erda.}
\end{figure}

\subsection{The Synergization of RDMA and NVM}

In recent years, synergizing RDMA and NVM  has become popular and important in order to obtain the salient features in these technologies. However, RDMA hardware does not support persistence guarantee for one-sided RDMA writes to NVM~\cite{yang2019orion}.  Therefore, if some data are transferred directly to remote NVM but part of the data are lost in the volatile NIC cache due to a failure, the data become incomplete and invalid.

Currently, providing persistence guarantee typically requires CPU participation or extra network round-trips~\cite{ChetDouglas_RDMA_with_PM}. We strive for  guaranteeing the persistence and atomicity of remote direct access with low overheads.

\section{Design}\label{sec:design}

\subsection{Erda Overview}

Erda is a zero-copy log-structured memory design that supports write-optimized Remote Data Atomicity (RDA) under RDMA and NVM scenarios. Figure~\ref{fig:architecture} shows the overall architecture of Erda. Specifically, data and metadata are persistent in a server's NVM. Data are stored in a log-structured manner following an array of head nodes. The append-only log always maintains an ¡°old¡± version of the updated data. Furthermore, a built-in checksum is used to verify the integrity of data without client-server coordination. Metadata stored in a hash table are used to index the data. We adopt a flexible flip bit and an $8$-byte atomic write in metadata to avoid redundant NVM writes as well as guarantee the atomicity of metadata. Clients perform read/write requests to the server using RDMA networking. Clients directly write data to the destination address (the log region) at servers without buffer and copy. In the following, we respectively present the structures of data and metadata, the access workflow, the write-optimized design, consistency guarantee, read-write competition and a log cleaning scheme in details.

\begin{figure}[t]
  \centering
    \includegraphics [width=0.48\textwidth]{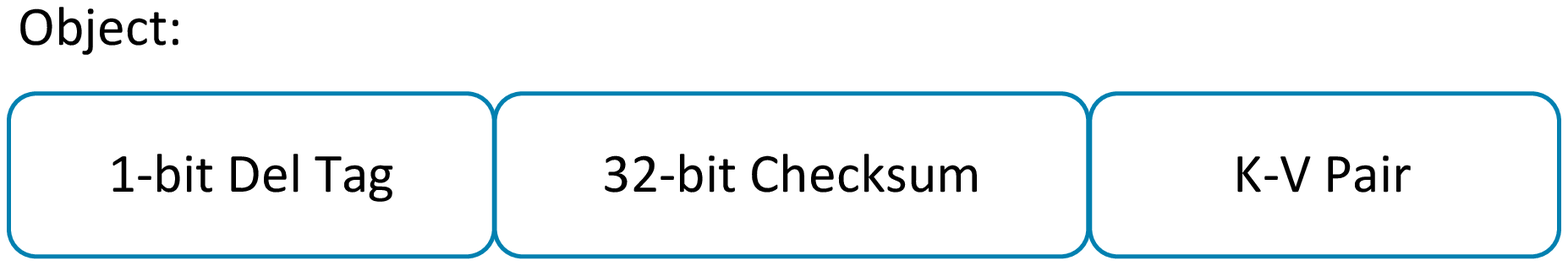}
\vspace{-5px}
    \caption{\label{fig:object} The object structure.}
\end{figure}

\begin{figure}[t]
\vspace{6px}	
  \centering
    \includegraphics [width=0.48\textwidth]{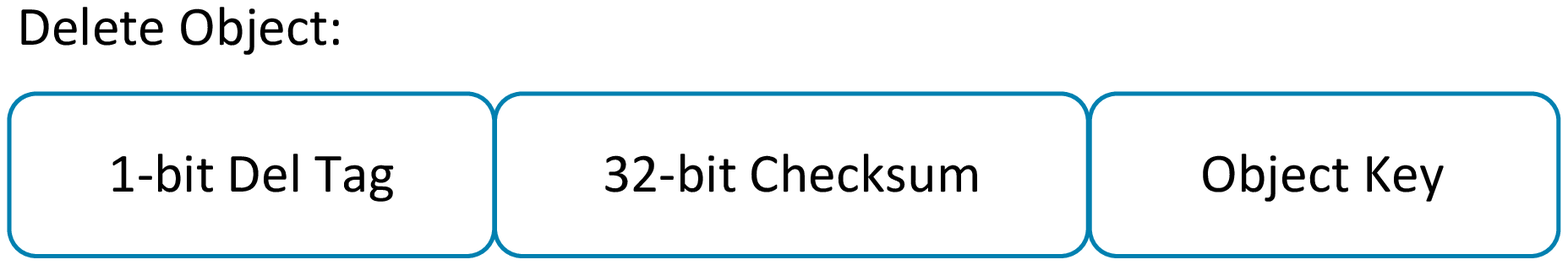}
\vspace{-5px}
    \caption{\label{fig:deletedObject} The structure of the deleted object.}
 \vspace{2px}
\end{figure}

\begin{figure*}[!ht]
  \centering
    \includegraphics [width=0.96\textwidth]{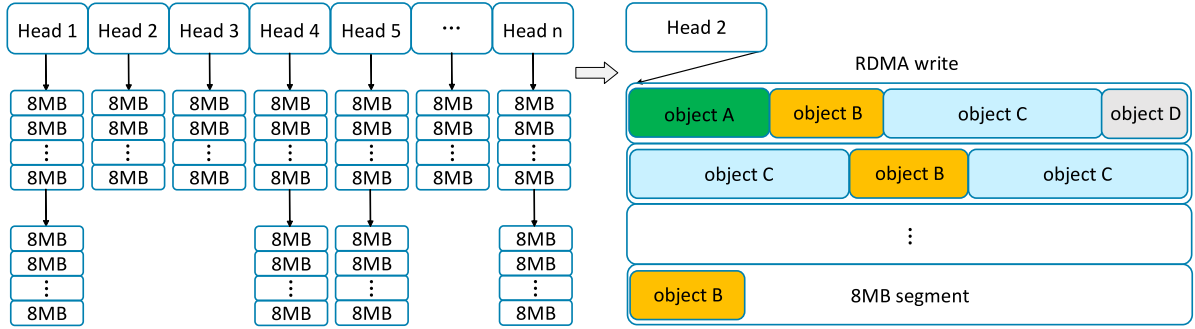}
\vspace{-5px}
    \caption{\label{fig:dataStructure} The structure of a log region where objects are stored.}
\end{figure*}

\subsection{The Structures of Data and Metadata}

\subsubsection{The Structure of a Normal/Deleted Object}

An object is the basic unit of one access and can be interpreted as a key-value pair with a checksum. As shown in Figure~\ref{fig:object}, an object consists of $1$-bit delete tag, $32$-bit CRC checksum and a key-value pair. Specifically, the delete tag indicates whether it is a normal object or a deleted one, which is shown in Figure~\ref{fig:deletedObject}. The $32$-bit CRC checksum computed over the entire object is used to check the integrity and the validity of the object. The last data field stores the key-value pair.

As shown in Figure~\ref{fig:deletedObject}, since Erda is a log-structured approach which appends all updates in an append-only log, we need the structure of the deleted object to indicate that the object has been deleted. The deleted object consists of $1$-bit delete tag (whose value is equal to $1$), $32$-bit CRC checksum and the object key. We do not need to store the value in the structure of the deleted object, which also saves storage space.

\subsubsection{The Structure of a Log Region}\label{subsubsec:PM Region}

\begin{figure}[t]
\vspace{3px}	
  \centering
    \includegraphics [width=0.48\textwidth]{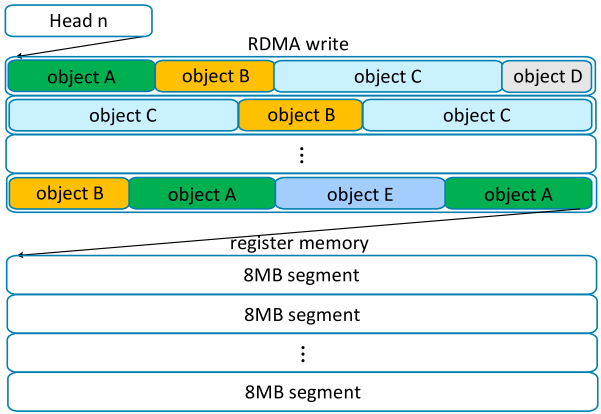}
\vspace{-5px}
    \caption{\label{fig:put7} Register memory for scalability.}
\end{figure}

We store and manage objects in each server using a log-structured manner. Figure~\ref{fig:dataStructure} shows the structure of a log region where objects are stored. Specifically, we use a head array of fixed addresses to link the log data, and the Head ID is used to distinguish different head nodes. Each head links a continuous memory region (such as $1$GB), and the continuous region is divided into $8$MB segments. For scalability, when a larger memory region is needed, we allocate and register another continuous $1$GB memory region and link it to the first $1$GB memory region following the same head, as shown in Figure~\ref{fig:put7}.

\begin{figure}[t]
\vspace{4px}	
  \centering
    \includegraphics [width=0.48\textwidth]{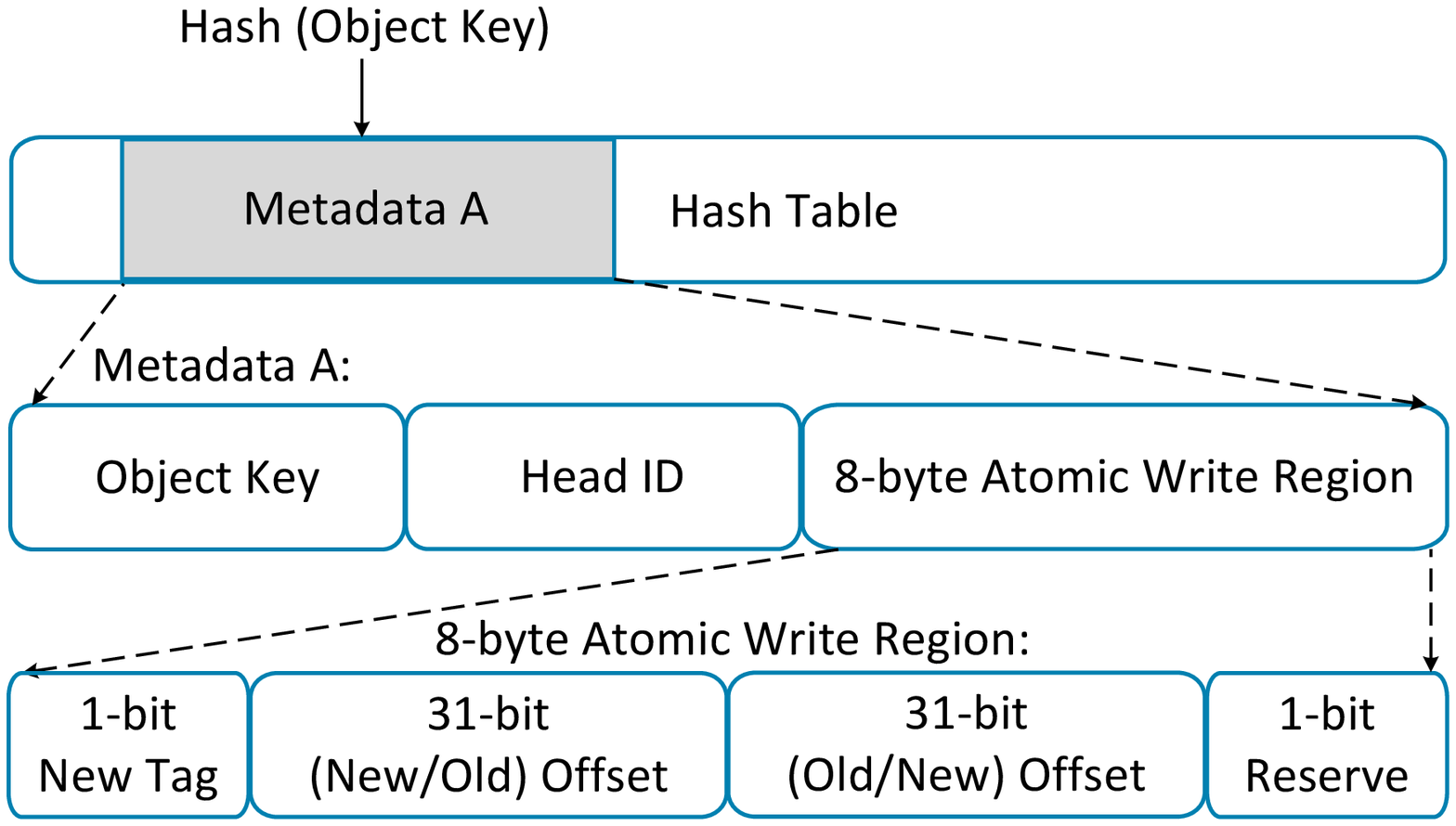}
\vspace{-5px}
    \caption{\label{fig:metadata} The metadata in a hash table.}
\end{figure}

\subsubsection{Metadata in a Hash Table}\label{subsubsec:Metadata}
We adopt flat namespace in a hash table to lookup objects. As shown in Figure~\ref{fig:metadata}, the entries in the hash table store the metadata of objects. An entry corresponding to an object stores the object key, the head ID and an $8$-byte atomic write region, including $1$-bit new tag which indicates whether the following $31$-bit data are ``new'' (the latest address information of the corresponding object) or ``old'' (the previous address information of the object), $31$-bit new/old offset, $31$-bit old/new offset and $1$-bit reserved position for future use. All the information in this region is updated in an $8$-byte atomic write.

\begin{figure*}[t]
  \centering
    \includegraphics [width=0.96\textwidth]{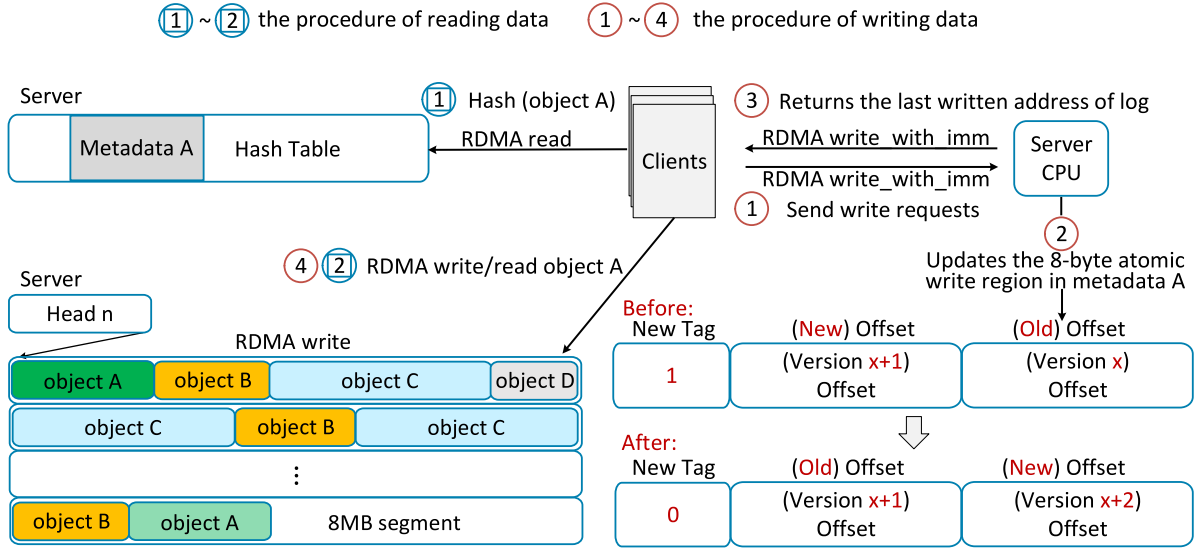}
\vspace{-5px}
    \caption{\label{fig:dataAccessMode} The procedures of reading and writing objects using one-sided RDMA.}
\end{figure*}

\subsection{Data Access Mode using RDMA}\label{subsec:Data Access}
Figure~\ref{fig:dataAccessMode} shows the procedures of reading and writing data (objects) using one-sided RDMA. To allow RDMA operations from a client, the server registers the memory regions of the metadata hash table and the log regions with RNIC (RDMA enabled NIC). Subsequently, with the corresponding remote registration keys, the client can issue RDMA operations to these memory regions. It is worth noting that once the connection is established, the server will send the head array containing the corresponding relationships between head IDs and pointers to the client.

We first describe the procedure of RDMA reads from the client to the server. After a client and a server establish a connection, according to the requested object key, the client uses an RDMA read to directly read the corresponding entry of the hash table in the server. Then, after verifying the received object key, the client queries the local cached head array for the pointer corresponding to the received head ID. Finally, with the aid of the $8$-byte atomic write region and the pointer, the client directly fetches the requested object using an RDMA read. When the client verifies the checksum of the object correctly, this RDMA read operation finishes.

For the procedure of RDMA writes from the client to the server, the client sends a write request to the server using RDMA write\_with\_imm, where the client's identifier is attached in the immediate data field. Moreover, the server updates the corresponding entry of a hash table and then returns the last written address of the log that is maintained and updated by the server. With the returned information, the client posts one-sided RDMA writes to directly write data in the log region of the remote server without participation of the server's CPU, and the server obtains higher processing capacity and removes redundant copy.

In the log region, the object does not span two segments.  When an object exceeds the current $8$MB segment, the server will change the last written address of the log to the beginning of the next $8$MB segment. For scalability, as described in Section~\ref{subsubsec:PM Region}, when a larger memory region is needed, we allocate and register another continuous $1$GB memory region and link it to the first $1$ GB memory region following the same head.

\section{Implementation Details}\label{sec:implementation}

\subsection{Write-Optimized Design for NVM}\label{subsec:Write-Optimized}

The write-optimized design consists of two components:

\textbf{Zero-Copy Memory Design.} We implement a zero-copy log-structured memory design. All the data are transferred directly from the clients to the log region at servers via RDMA writes, and due to out-of-place updates, we do not need to put the data into some buffers like redo logging. However, this zero-copy design may bring some consistency issue such as partial write. The corresponding unique consistency detection and recovery are shown in Section~\ref{subsec:Consistency}.

\textbf{Flexible Flip Bit.} We adopt a flip bit, named ``New Tag'', in a hash table to indicate whether the next region is the new or old offset,  thus avoiding redundant NVM writes. When a server receives an update request, it locates the hash entry according to the hash value of the requested object key. The modification of the hash entry consists of two steps. First, flip the ``New Tag''. Second, write the last written address of the log (i.e., the offset) to one of the $31$-bit regions according to the ``New Tag'' to be written. If the ``New Tag'' to be written is $1$, write the address to the first $31$-bit region; otherwise, write the address to the second $31$-bit region. As shown in the lower right part of Figure~\ref{fig:dataAccessMode}, the ``New Tag'' in the $8$-byte atomic write region of metadata A is $1$ before being updated. Then, the server flips the ``New Tag'' from $1$ to $0$ and writes the address to the second $31$-bit region. The part with unchanged contents will skip bit programming action and not be written using data-comparison write (DCW)~\cite{yang2007low}.

\subsection{Consistency Detection and Recovery}\label{subsec:Consistency}

Erda is able to support consistency and atomicity of RDMA operations:

\textbf{Out-of-Place Updates.} We adopt a log-structured memory to prevent in-place updates, and always maintain an ``old'' version of the updated object (similar to an undo log).

\textbf{CRC Checksum.} We add a $32$-bit CRC checksum over each object, so clients can detect the incompleteness of the fetched object by verifying the checksum.  Once the fetched object is incomplete, the client can issue another RDMA read to fetch the previous version of the object.

\textbf{$8$-Byte Atomic Write.}  We leverage an $8$-byte atomic write in the entry of a hash table, so the inconsistency will only occur when the metadata in the entry of a hash table have been atomically updated but a failure occurs before the object data have been fully written into the log. At the same time, the $8$-byte atomic write also contains the address information for the ``old'' object version. When a failure occurs, the server can properly restore to a consistent ``old'' version.

\begin{figure}[t]
  \centering
    \includegraphics [width=0.48\textwidth]{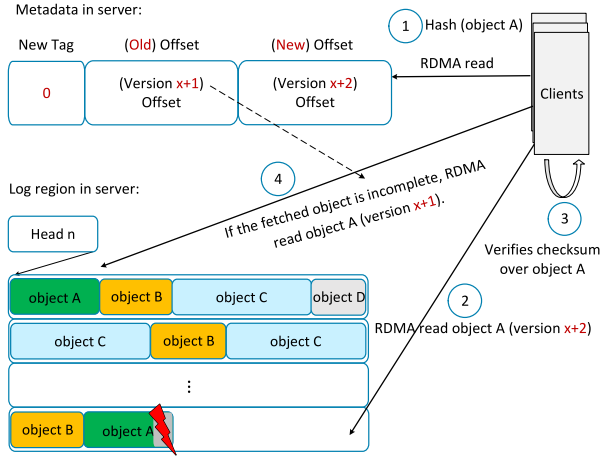}
    \caption{\label{fig:crash} If a failure occurs during a previous RDMA write operation, other clients detect the inconsistency when they access the incomplete object. These clients obtain a previous consistent version. }
\end{figure}

For example, if a client fails when it is updating object $A$ using RDMA write, the latest object $A$ in server's log is incomplete. However, the server is unaware of the incomplete object due to the one-sided RDMA operation without involving the server's CPU. As shown in Figure~\ref{fig:crash}, if a client accesses object $A$ later, it will realize that the fetched object $A$ is incomplete by verifying the checksum. Then, the client can issue another RDMA read to fetch the previous version of object $A$ based on the old offset in the $8$-byte atomic write region which has been already fetched. The client will also inform the server to update the corresponding entry in a hash table (replace the current new offset with the old offset). Thus, all subsequent accesses to the current object $A$ will be correct. Furthermore, once a failure in a server results in incomplete objects, the server needs to check objects in the last segment following each head and correspondingly update metadata in the hash table for consistency.

\subsection{Read-Write Competition}\label{subsec:Read-Write Races}

After a client sends a write request to a server, the server will reserve the corresponding object storage region in the $1$ GB continuous memory region and update the last written address of the log. When the server receives another write request to write an object following the same head node, the server will return the updated last written address. A specific storage region of an object is only be written by one client, but each memory region is read by many clients. Thus, there is no write-write competition. However, the RDMA write may create read-write competition with concurrent RDMA reads by other clients.

When performing a write operation, the modification in the entry of a hash table is an atomic operation as described in Section~\ref{subsec:Consistency}, and there are two read-write scenarios. First, when the server has modified a entry of a hash table atomically after receiving a write request from a client, but the client has not completed the object write, synchronous read operations from other clients find that the object is invalid for read by checking the checksum, or the object is a null value since the object being read has not been written yet. In these cases, the clients for reading choose to read a previous version of the requested object by using the old offset from the obtained entry of a hash table, or just wait a moment and try to read the same address again. Second, when a client has read the entry but not read the corresponding object, another client modifies the same entry and writes the updated object in the log at the same time. The read-write competition in this case does not lead to errors, because the update in our log-structured mechanism is an out-of-place update.

\begin{figure}[t]
  \centering
    \includegraphics [width=0.48\textwidth]{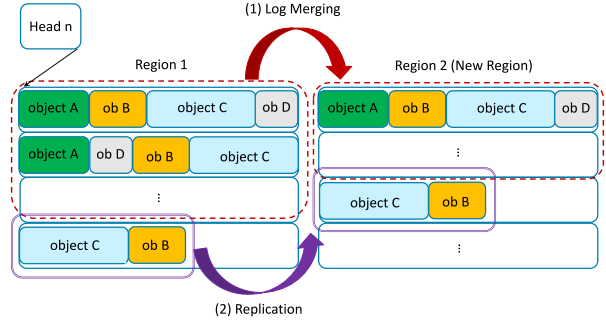}
\vspace{-5px}
    \caption{\label{fig:LogCleaning1} Log cleaning consists of two phases: log merging and replication.}
\end{figure}

\subsection{Lock-Free Log Cleaning}\label{subsec:Log Cleaning}

Log cleaning reclaims free space of the append-only log by removing deleted objects and stale versions of objects for memory saving. A server performs log cleaning and handles read/write requests concurrently. As shown in Figure~\ref{fig:LogCleaning1}, log cleaning consists of two phases: log merging and replication. We use an example to illustrate the process of log cleaning.

\begin{figure}[t]
  \centering
    \includegraphics [width=0.48\textwidth]{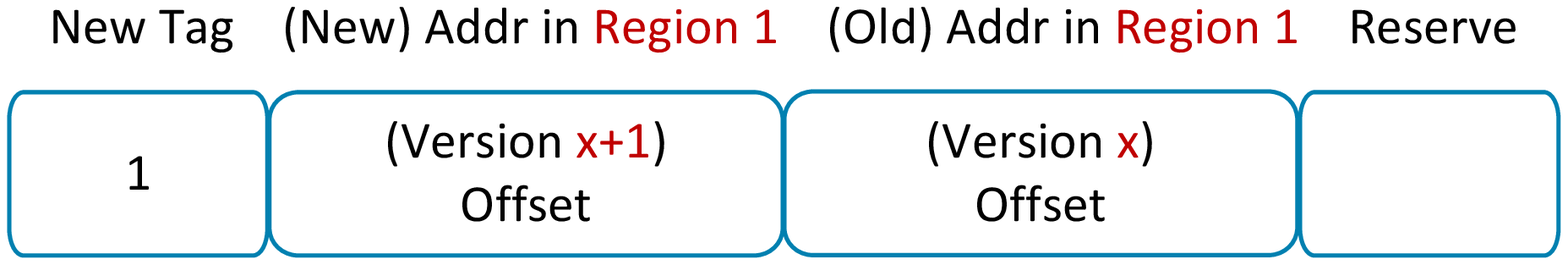}
\vspace{-7px}
    \caption{\label{fig:Cleaning1} The $8$-byte atomic write region in metadata before log cleaning.}
\end{figure}

\begin{figure}[t]
  \centering
    \includegraphics [width=0.48\textwidth]{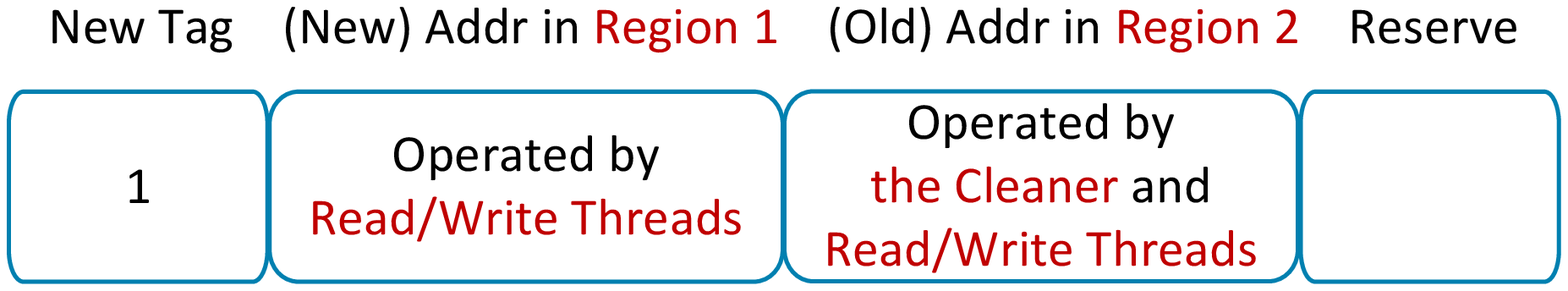}
\vspace{-7px}
    \caption{\label{fig:Cleaning2} The $8$-byte atomic write region in metadata during the log cleaning.}
\end{figure}

\begin{figure}[t]
  \centering
    \includegraphics [width=0.48\textwidth]{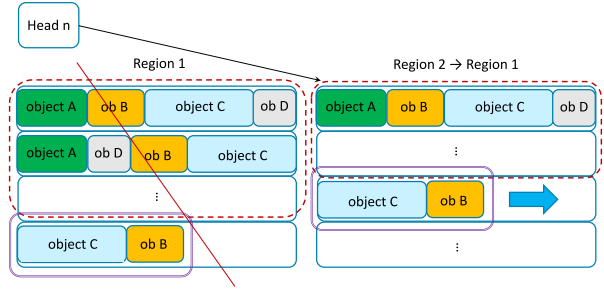}
\vspace{-5px}
    \caption{\label{fig:LogCleaning2} After completing log cleaning, Region $2$ becomes Region $1$.}
\end{figure}

\begin{figure}[t]
  \centering
    \includegraphics [width=0.48\textwidth]{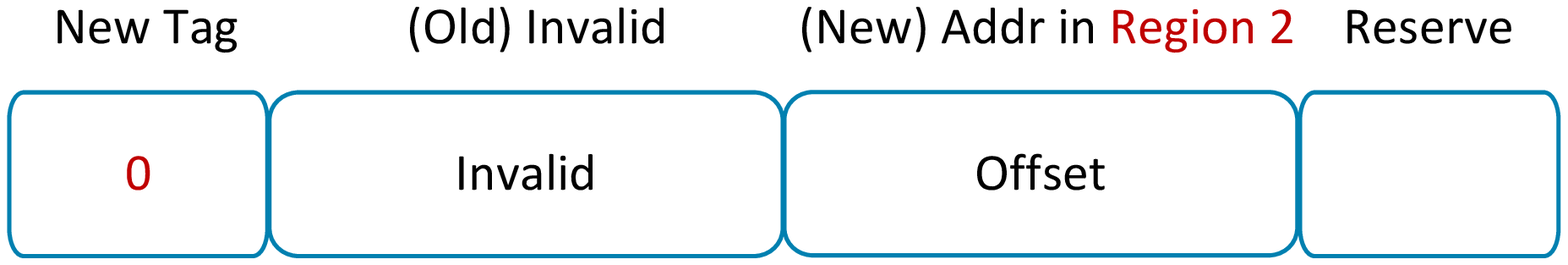}
\vspace{-7px}
    \caption{\label{fig:Cleaning3}  When a log cleaning process is completed, the server flips the new tag from $1$ to $0$, which means the address information in Region $2$ is the latest version and will be accessed by clients.}
\end{figure}
When the occupied space following a head reaches a pre-defined threshold (Region $1$ in Figure~\ref{fig:LogCleaning1}) , the cleaner in a server will allocate another continuous $1$ GB memory region (Region $2$ in Figure~\ref{fig:LogCleaning1}), as well as inform all the connected clients that the objects following the head will experience log cleaning. After receiving the notification, clients can still read and write objects, but in different ways: clients issue read/write requests using RDMA send. Furthermore, in the $8$-byte atomic write region of metadata, the server doesn't flip the new tag. Based on the new tag, the previous ``new offset region'' now stores the address information of Region $1$, and the ``old offset region'' stores the address information of Region $2$, as shown in Figures~\ref{fig:Cleaning1} and~\ref{fig:Cleaning2}. At the same time, the cleaner in the server starts log cleaning after going through maximum RTT and informing connected clients to avoid transmission delays. The cleaner also doesn't flip the new tag, and merely updates the old offset region.

In the log merging phase, the cleaner starts the reverse scan from the last written address of the log at the beginning of log cleaning, since the object version of the later part of the log is newer than that of the previous part. For the object that is first encountered (representing the latest version in the merging region), the cleaner writes it to Region $2$ and updates the corresponding old offset region in the entry. Furthermore, when the cleaner encounters the same object (the stale version) again, it simply overlooks it. In addition, the deleted objects will be removed during the cleaning process. For read/write requests from clients, the server accesses the new offset region in Region $1$.

When the reverse scan is completed, the log cleaning moves on to the replication phase. The cleaner in a server replicates objects that were written by clients after the start of the log merging phase into Region $2$, and the server handles read/write requests concurrently. Specifically, for the write requests from clients, the server updates the old offset region in Region $2$ in the entry and appends the new object into Region $2$. The replication region in Region $2$ is reserved for the cleaner. If the object to be replicated has already appeared in the following written region, the entry (the old offset region in Region $2$) will not be changed, since the offset is  the latest version. For the read requests from clients, if the offset in Region $2$ in the corresponding entry is larger than the offset at the end of the reserved replication region, the server reads the address in the old offset region in Region $2$ (the latest version); otherwise, the server reads the address in the new offset region in Region $1$, since some data in Region $1$ fail to be replicated into Region $2$.

When all the objects that were written after the start of the log merging phase in Region $1$ are replicated into Region $2$, the log cleaning process is completed. At this point, the server changes the pointer of the corresponding head from pointing to Region $1$ to Region $2$, as shown in Figure~\ref{fig:LogCleaning2}. Then, the server flips the new tags in the hash tables of all the objects in Region $2$ (Figure~\ref{fig:Cleaning3}), returns the new pointer to the connected clients and informs these clients that the log cleaning is finished. After that, clients return to the original ways of reading and writing objects.

\section{Performance Evaluation and Analysis}\label{sec:evaluation}

We examine the performance of Erda in terms of multiple metrics, including the number of NVM writes, throughput and latency.

\begin{figure}[t]
  \centering
    \includegraphics [width=0.48\textwidth]{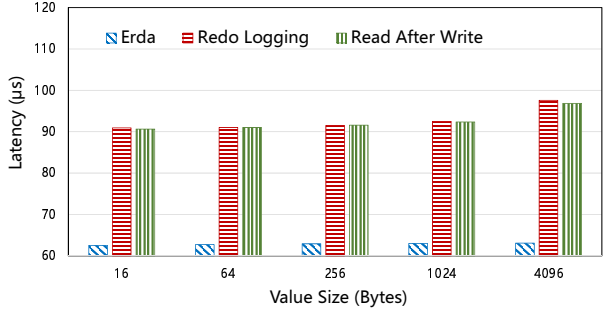}
\vspace{-7px}
    \caption{\label{fig:LatencyYCSB-C} The latency of YCSB-C (100\% read) with different value sizes of the key-value pair.}
\end{figure}

\begin{figure}[t]
  \centering
    \includegraphics [width=0.48\textwidth]{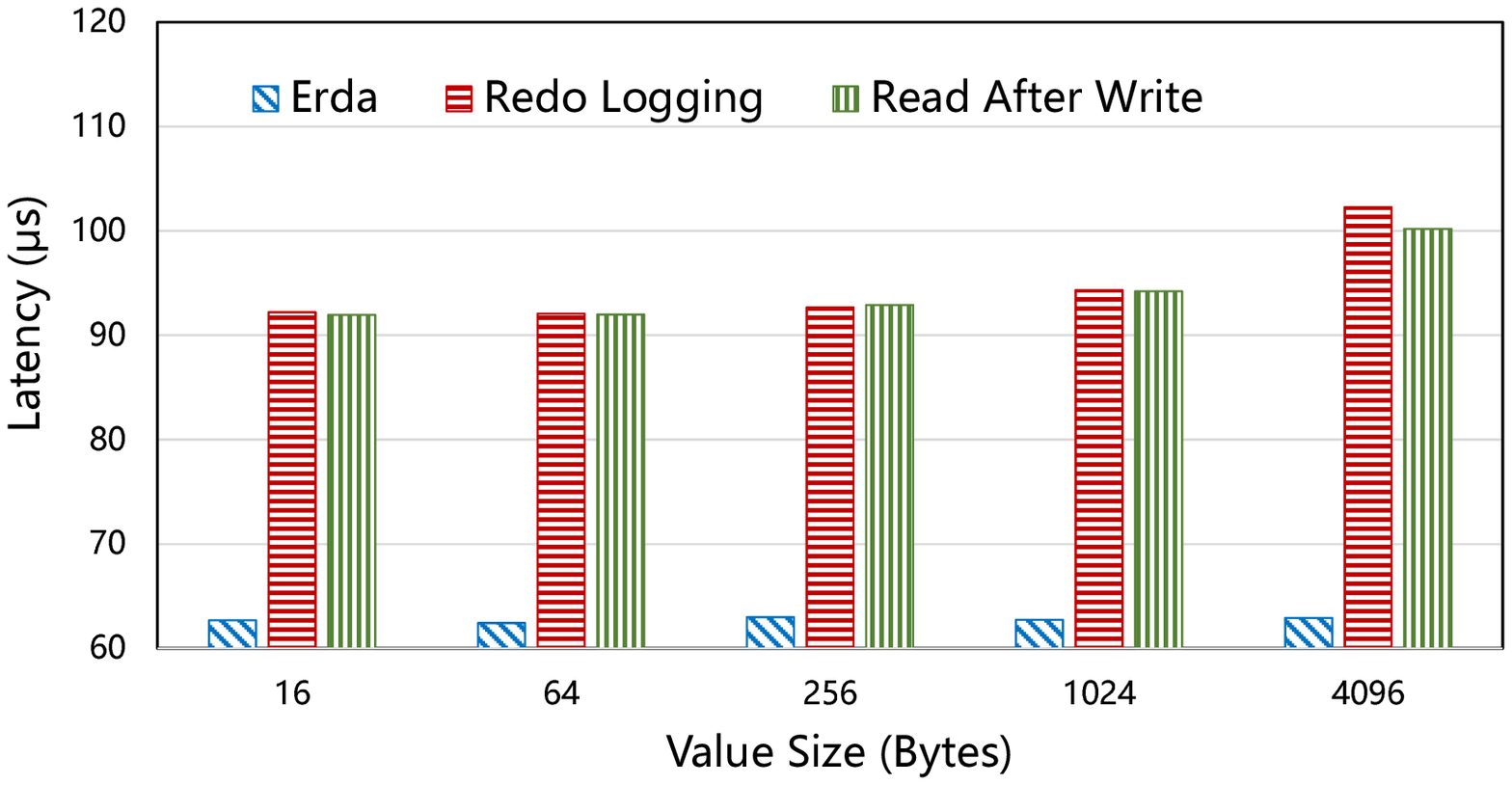}
\vspace{-7px}
    \caption{\label{fig:LatencyYCSB-B} The latency of YCSB-B (95\% read, 5\% write) with different value sizes of the key-value pair.}
\end{figure}

\begin{figure}[t]
  \centering
    \includegraphics [width=0.48\textwidth]{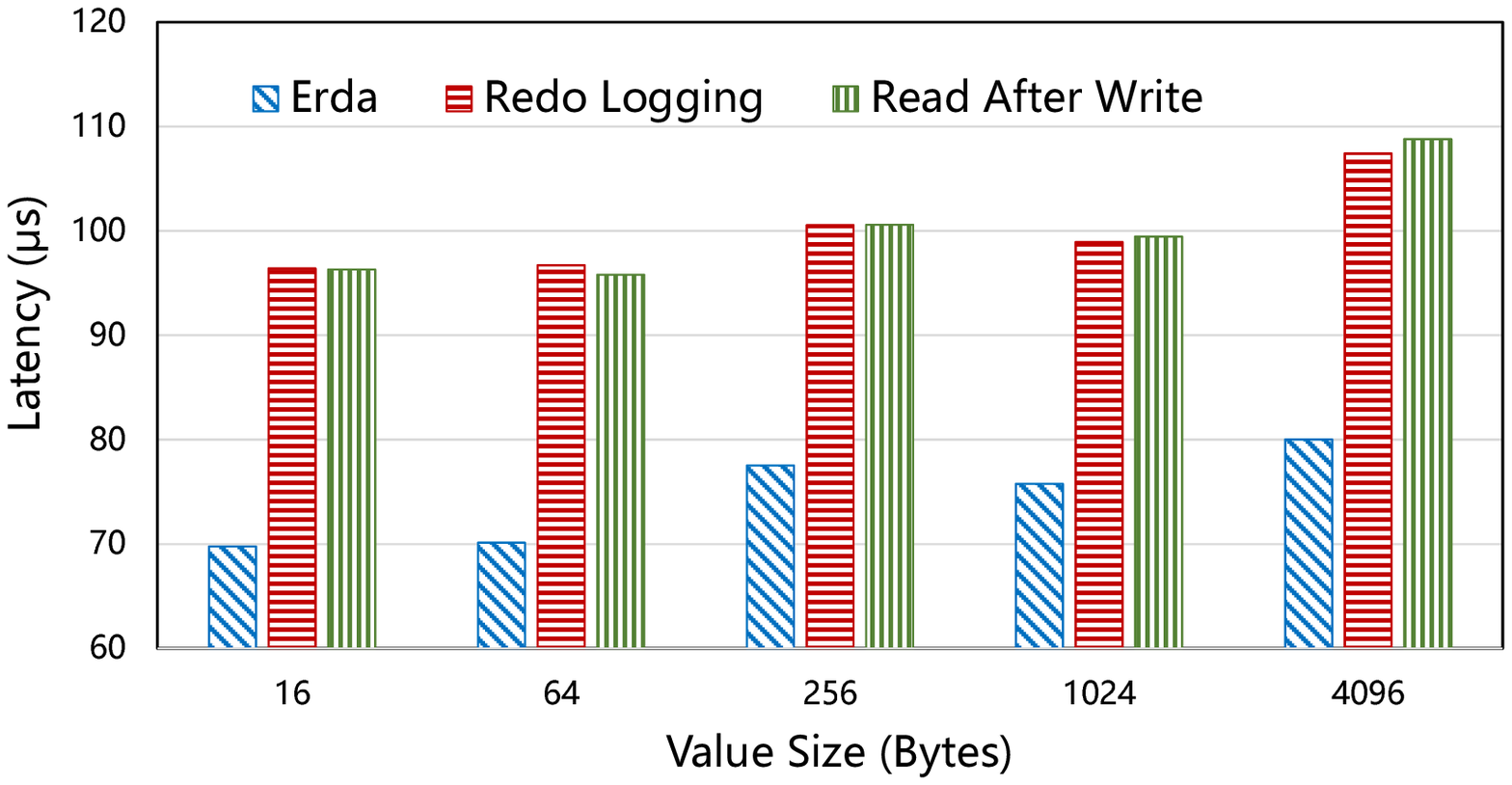}
\vspace{-7px}
    \caption{\label{fig:LatencyYCSB-A} The latency of YCSB-A (50\% read, 50\% write) with different value sizes of the key-value pair.}
\end{figure}

\begin{figure}[t]
  \centering
    \includegraphics [width=0.48\textwidth]{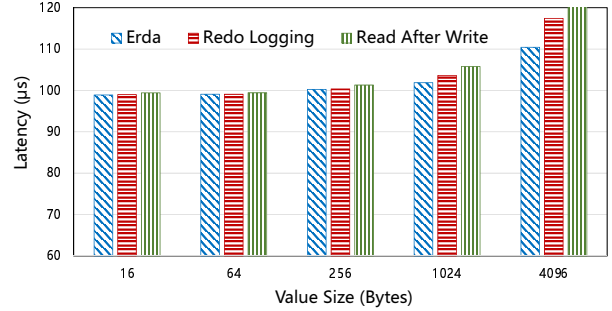}
\vspace{-7px}
    \caption{\label{fig:Latency100Put} The latency of update-only workload (100\% write) with different value sizes of the key-value pair.}
\end{figure}

\subsection{Experimental Setup}

\textbf{Hardware and configurations.} Our experiments run upon the servers, each of which contains two $2.4$ GHz Intel Xeon $E5620$ CPUs ($4$ cores) and $12$GB of DDR$3$ RAM. One server is also equipped with a $40$Gbps Mellanox ConnectX-3 InfiniBand network adapter and runs on CentOS $7.3$ with the MLNX\_OFED\_LINUX-$4.3$ InfiniBand driver. As real NVM devices are not fully available, we adopt a well-recognized simulation method that adds extra write latency of DRAM to simulate NVM  ~\cite{hu2017log,xia2017hikv,ou2016high,huang2014nvram,volos2011mnemosyne}. By default, we add $150$ns of extra write latencies~\cite{volos2011mnemosyne}.

\textbf{Workloads.} We use the YCSB benchmark~\cite{cooper2010benchmarking} to generate four workloads that follow Zipfian distribution of skewness $0.99$: ($1$) Read-only workload (YCSB-C) contains 100\% read. ($2$) Read-mostly workload (YCSB-B) contains 95\% read and 5\% write. ($3$) Update-heavy workload (YCSB-A) contains 50\% read and 50\% write. ($4$) Update-only workload contains 100\% write.

\textbf{Comparisons.} We compare Erda with two consistency schemes: Redo Logging  (a CPU involvement scheme)~\cite{ogleari2018steal,nguyen2018picl} and Read After Write (a network-dominant scheme)~\cite{ChetDouglas_RDMA_with_PM,ChetDouglas_Intel_Perspective}. For \textbf{Redo Logging} scheme, a client sends a write request to the redo log region of the server using RDMA send, and then the server verifies the integrity of the message in the redo log and applies the write request asynchronously to the destination storage. When a client issues an RDMA send to request an object value, the server first looks for the object in the redo log. If the requested object isn't in the redo log, the server searches the destination address with the object key through a hash table, and then reads the object and returns it to the client. For \textbf{Read After Write} scheme, to write objects, a client first sends a request to a server and obtains the address to be written in the ring buffers. Moreover, the client uses RDMA write to push the object into the ring buffers, and issues an RDMA read following the RDMA write to force the object to be persistent and integrated into the ring buffers. The server CPU polls for these operations asynchronously from ring buffers and applies them to the destination storage. The procedure of read operations from the client to the server follows the operations of redo logging scheme.

Redo Logging, Read After Write and Erda leverage hopscotch hashing algorithm~\cite{HopscotchHashing} to index objects. In the hopscotch hashing, a key-value pair locates in a small contiguous region of memory, while in cuckoo hashing~\cite{CuckooHashing}, a key-value pair is in one of several disjoint regions~\cite{dragojevic2014farm}.

\subsection{Latency}
\begin{figure}[t]
  \centering
    \includegraphics [width=0.48\textwidth]{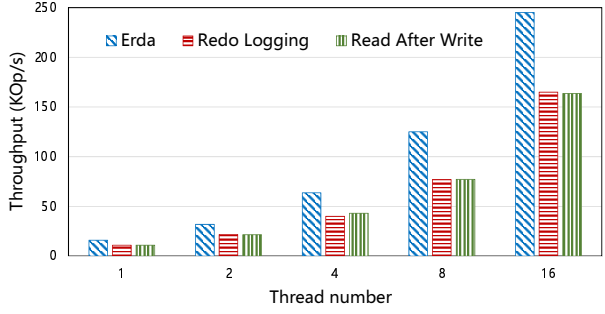}
\vspace{-7px}
    \caption{\label{fig:ThroughputYCSB-C} The throughput of YCSB-C (100\% read) with different thread numbers.}
  \vspace{-7px}
\end{figure}

\begin{figure}[t]
  \centering
    \includegraphics [width=0.48\textwidth]{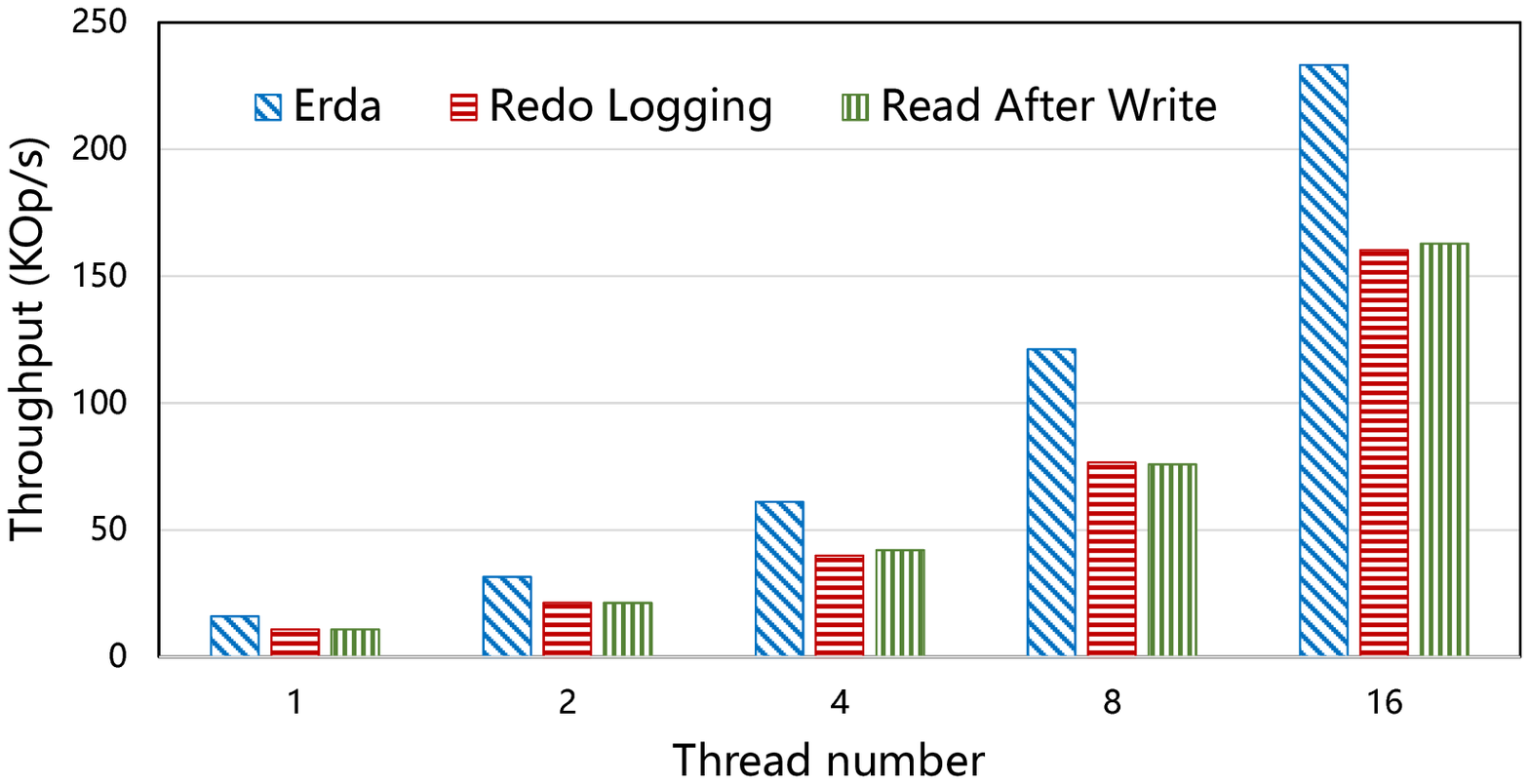}
\vspace{-7px}
    \caption{\label{fig:ThroughputYCSB-B} The throughput of YCSB-B (95\% read, 5\% write) with different thread numbers.}
  \vspace{-7px}
\end{figure}
\begin{figure}[t]
  \centering
    \includegraphics [width=0.48\textwidth]{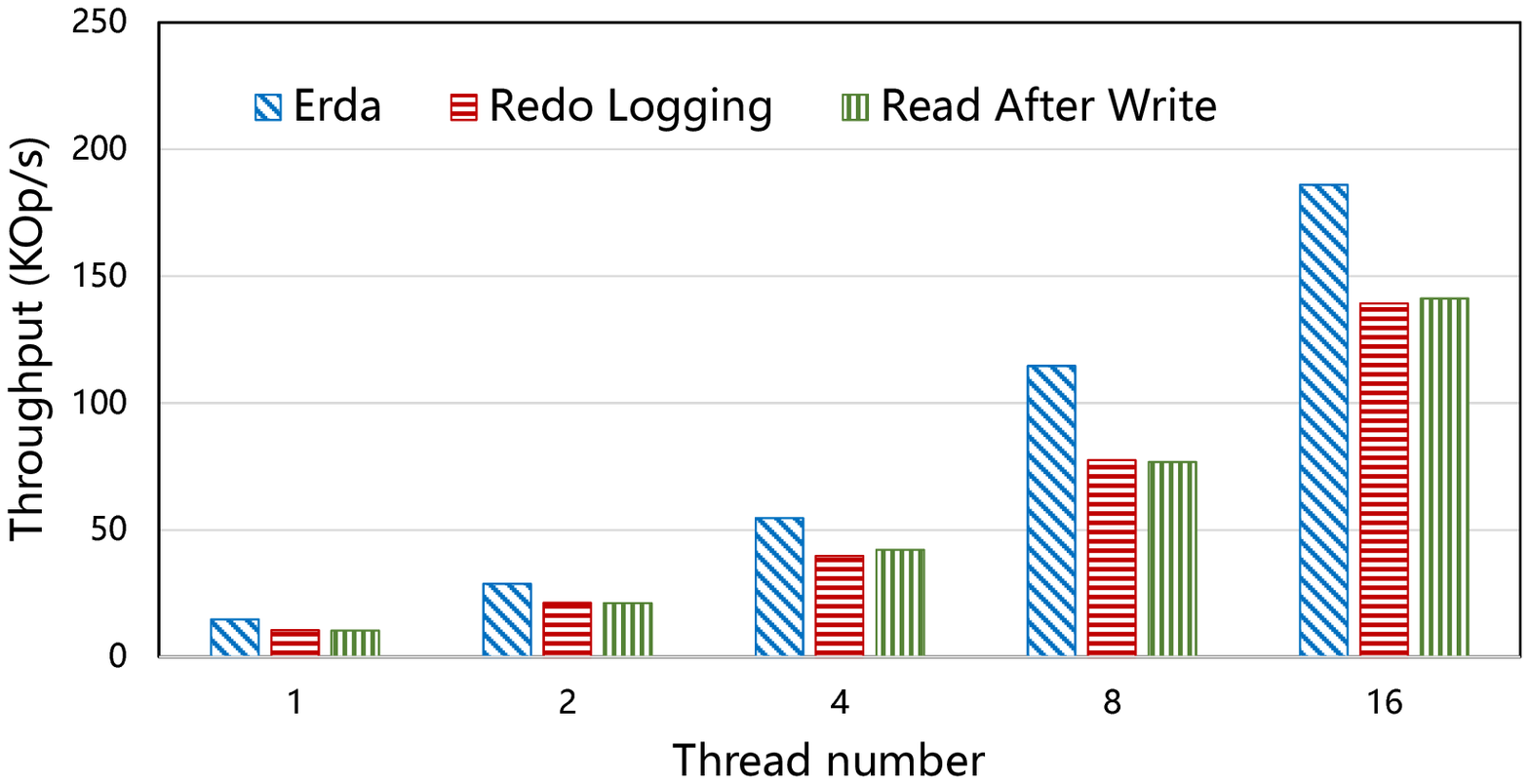}
\vspace{-7px}
    \caption{\label{fig:ThroughputYCSB-A} The throughput of YCSB-A (50\% read, 50\% write) with different thread numbers.}
  \vspace{-7px}
\end{figure}

\begin{figure}[t]
  \centering
    \includegraphics [width=0.48\textwidth]{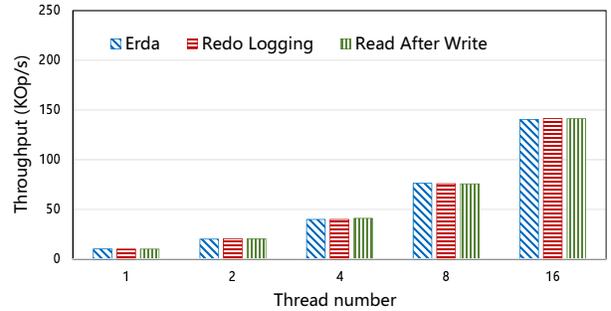}
\vspace{-7px}
    \caption{\label{fig:Throughput100Put} The throughput of update-only workload (100\% write) with different thread numbers.}
  \vspace{-7px}
\end{figure}
As shown in Figures~\ref{fig:LatencyYCSB-C} --~\ref{fig:Latency100Put}, we compare the latency of Erda with those of Redo Logging and Read After Write using four YCSB workloads, as the value size of the key-value pair varies from $16$Bytes to $4,096$Bytes. In Figures~\ref{fig:LatencyYCSB-C} and~\ref{fig:LatencyYCSB-B} where read operations dominate the workloads, Erda performs much better than Redo Logging and Read After Write. In Redo Logging and Read After Write, clients send read requests to a server using two-sided RDMA send. However, when receiving the requests, the server needs to first identify the requested object in the redo log. If the object isn't in the log, the server reads the object from the destination address, and then returns the object to clients. In Erda, clients use two one-sided RDMA reads to perform read operations (one for the corresponding entry of the hash table in the server, and the other for directly fetching the requested object) without the CPU involvements of servers. Specifically, the average latency of YCSB-C (100\% read) for Erda, Redo Logging and Read After Write are $62.84$$\mu$s, $92.7$$\mu$s and $92.48$$\mu$s, respectively. The average latencies of YCSB-B (95\% read, 5\% write) for Erda, Redo Logging and Read After Write are $62.76$$\mu$s, $94.71$$\mu$s and $94.25$$\mu$s, respectively. Figure~\ref{fig:LatencyYCSB-A} shows the latency of YCSB-A (50\% read, 50\% write) with different value sizes of the key-value pair. The corresponding average  latencies for Erda, Redo Logging and Read After Write are $74.64$$\mu$s, $100$$\mu$s and $100.18$$\mu$s, respectively. For update-only workload (100\% write) shown in Figure~\ref{fig:Latency100Put}, Erda still outperforms the other two schemes, although the benefits of using Erda with the update-only workload are less than that with other three workloads. Specifically, the average latencies of update-only workload for Erda, Redo Logging and Read After Write are $102.1$$\mu$s, $103.89$$\mu$s and $105.47$$\mu$s, respectively.

\subsection{Throughput}

Figures~\ref{fig:ThroughputYCSB-C} --~\ref{fig:Throughput100Put} show the throughputs of Erda, Redo Logging and Read After Write with four YCSB workloads and different thread numbers, respectively. From Figure~\ref{fig:ThroughputYCSB-C}, we observe that the throughput of Erda grows approximately linearly with the increasing thread numbers, while Redo Logging and Read After Write don't. The main reason is that YCSB-C is a read-only workload, and the read procedure of Erda from clients to servers does not involve server CPUs by using two one-sided RDMA reads, while the read procedures of Redo Logging and Read After Write need CPU involvements. Hence the throughput of Erda is not affected by CPU consumption as the number of threads increases. Specifically, the average throughputs of YCSB-C for Erda, Redo Logging and Read After Write are $96.35$KOp/s, $62.93$KOp/s and $63.28$KOp/s, respectively. As shown in Figure~\ref{fig:ThroughputYCSB-B}, the average throughput of YCSB-B for Erda, Redo Logging and Read After Write are $92.57$KOp/s, $61.78$KOp/s and $62.57$KOp/s, respectively. For YCSB-A workload shown in Figure~\ref{fig:ThroughputYCSB-A}, the average throughput of Erda, Redo Logging and Read After Write are $79.77$KOp/s, $57.60$KOp/s and $58.32$KOp/s, respectively. However, for update-only workload (100\% write) shown in Figure~\ref{fig:Throughput100Put}, the average throughputs of Erda, Redo Logging and Read After Write are approximate.

\subsection{CPU Utilization}

\begin{figure}[t]
  \centering
    \includegraphics [width=0.48\textwidth]{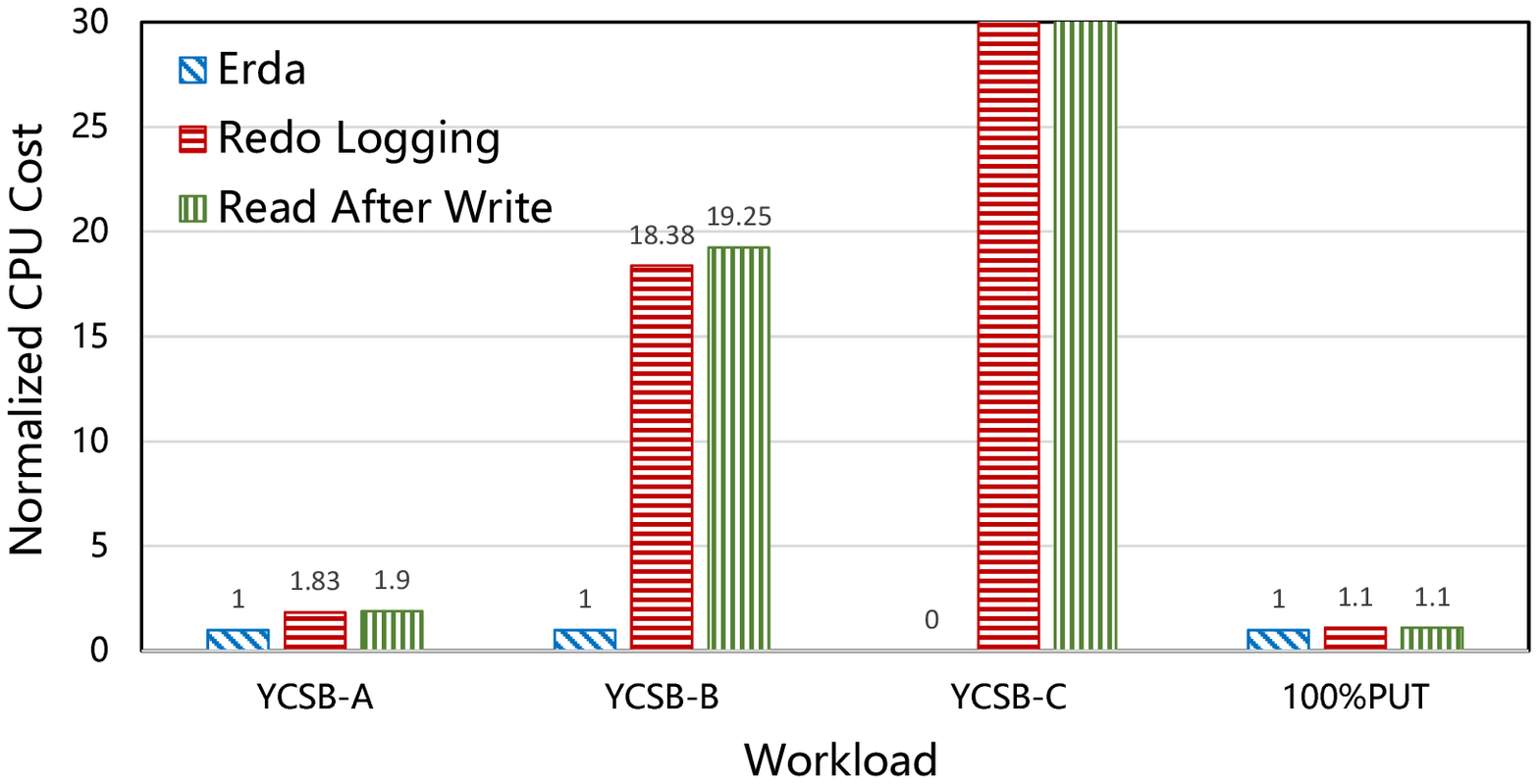}
\vspace{-7px}
    \caption{\label{fig:CPU16Bytes} The normalized CPU cost when the value size of the key-value pair is $16$Bytes.}
  \vspace{-7px}
\end{figure}

\begin{figure}[t]
  \centering
    \includegraphics [width=0.48\textwidth]{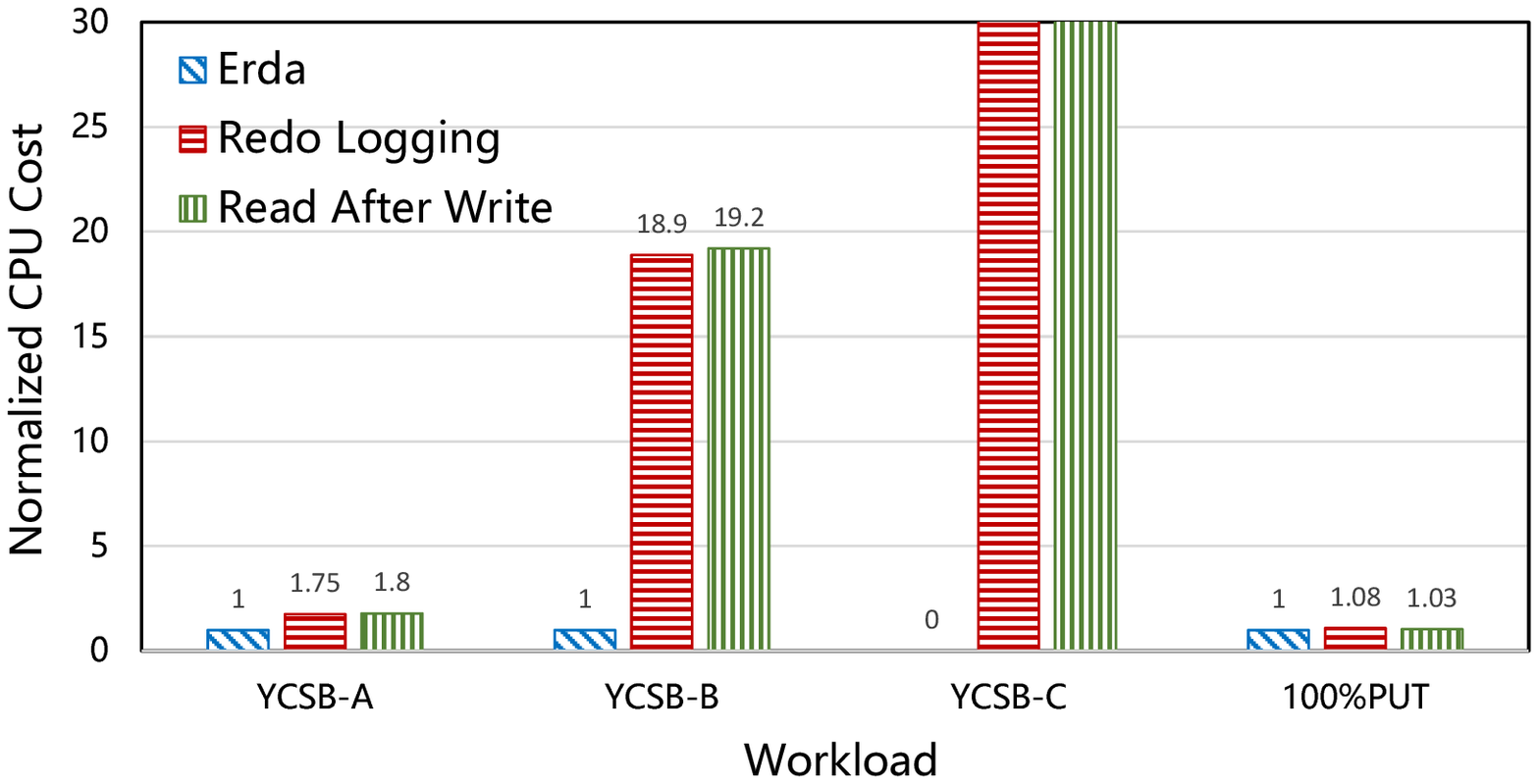}
\vspace{-7px}
    \caption{\label{fig:CPU64Bytes} The normalized CPU cost when the value size of the key-value pair is $64$Bytes.}
  \vspace{-7px}
\end{figure}

\begin{figure}[t]
  \centering
    \includegraphics [width=0.48\textwidth]{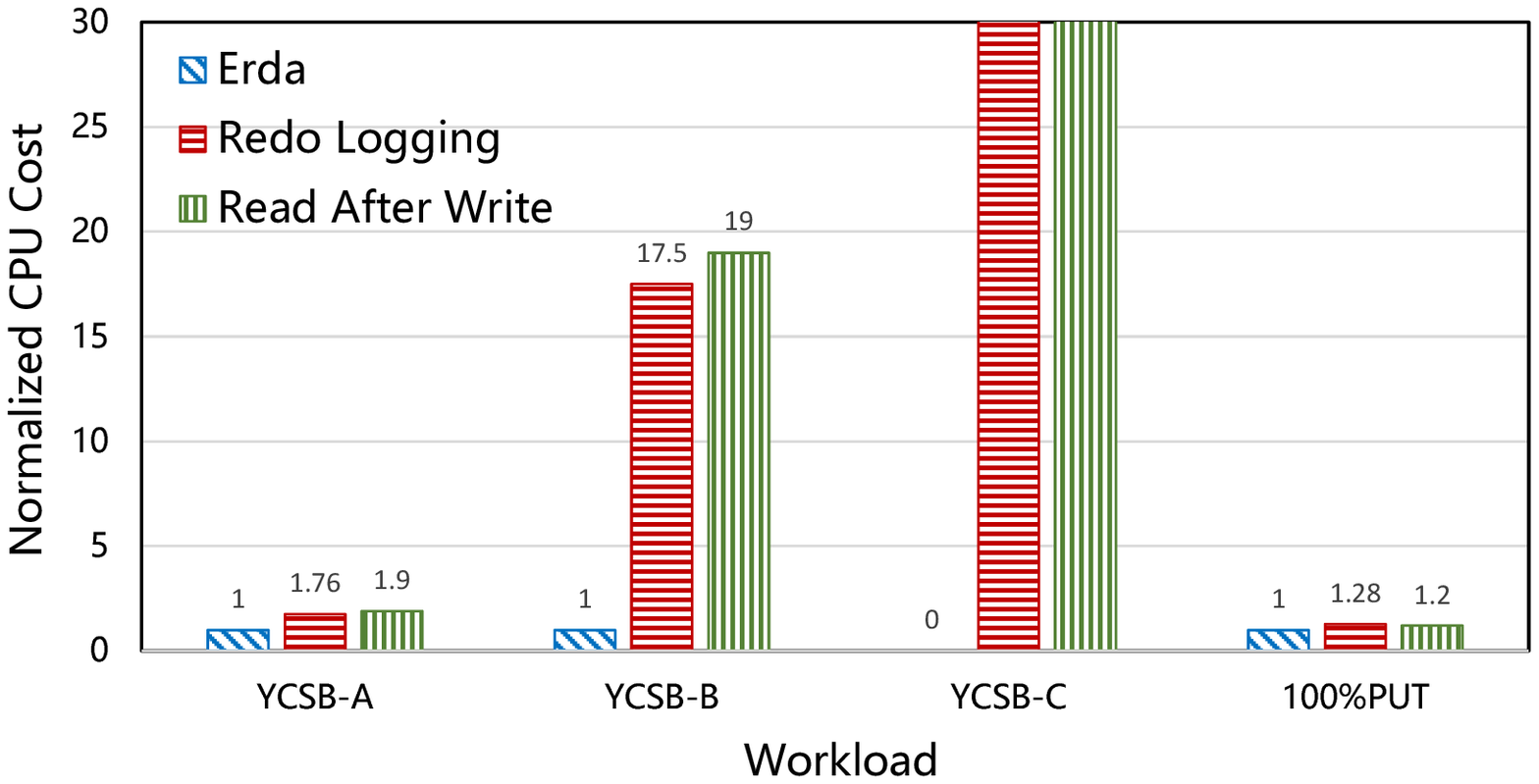}
\vspace{-7px}
    \caption{\label{fig:CPU256Bytes} The normalized CPU cost when the value size of the key-value pair is $256$Bytes.}
  \vspace{-7px}
\end{figure}

\begin{figure}[t]
  \centering
    \includegraphics [width=0.48\textwidth]{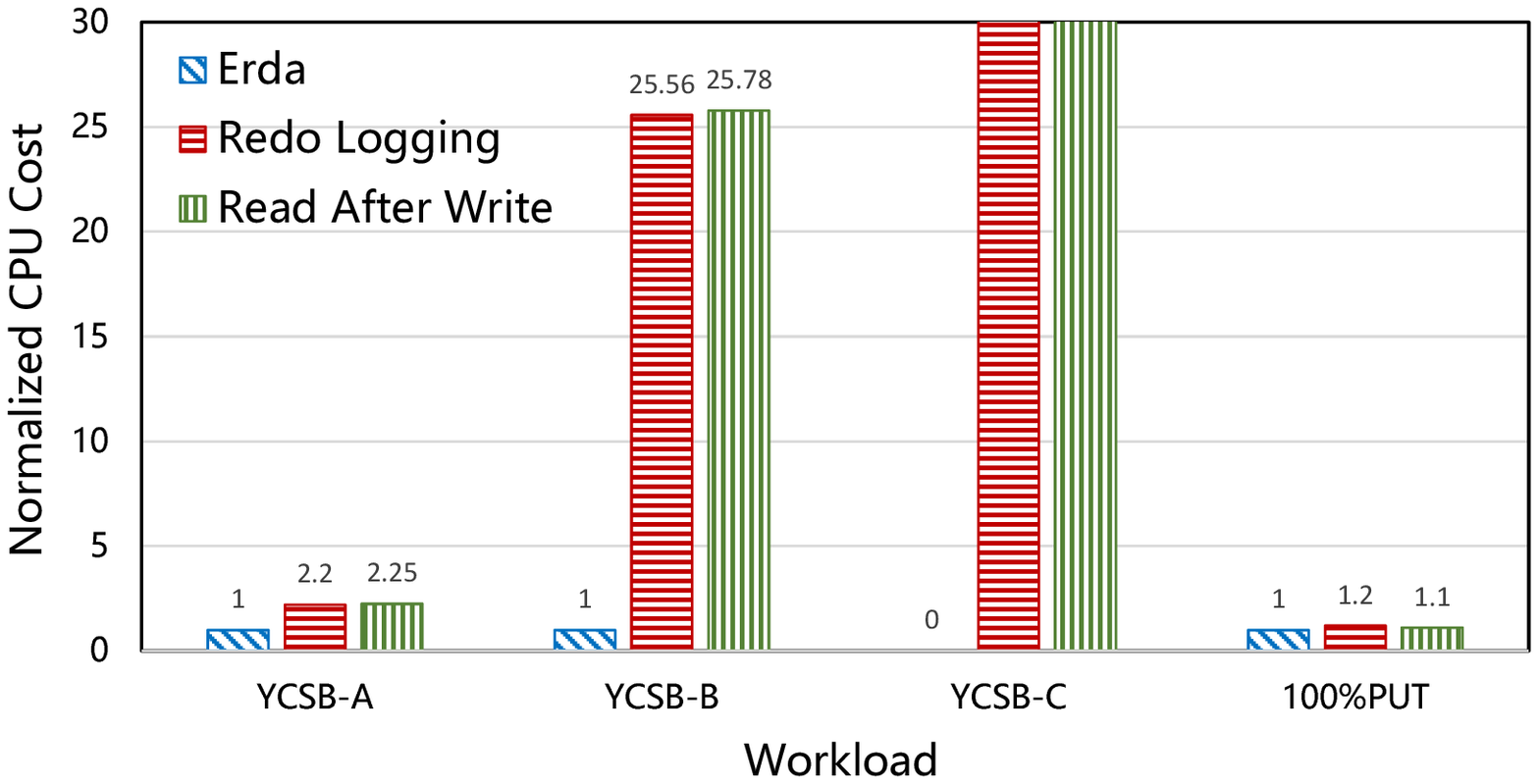}
\vspace{-7px}
    \caption{\label{fig:CPU1024Bytes} The normalized CPU cost when the value size of the key-value pair is $1024$Bytes.}
  \vspace{-7px}
\end{figure}

We use the ``top'' command in Linux to measure CPU utilization, and show the results of the normalized CPU costs with different workloads and value sizes of the key-value pair in Figures~\ref{fig:CPU16Bytes} --~\ref{fig:CPU1024Bytes}. For YCSB-C workload (100\% read), the CPU cost of Erda is $0$ since the read procedure of Erda does not involve server CPUs. Hence the normalized CPU costs of both Redo Logging and Read After Write are positive infinity. Due to the same reason, for YCSB-B workload (95\% read), the normalized CPU costs of both Redo Logging and Read After Write are much higher than that of Erda. Specifically, the normalized CPU costs of Redo Logging and Read After Write for YCSB-B workload are on average $20.09$ and $20.81$ times higher than the cost of Erda, respectively. For YCSB-A workload (50\% read and 50\% write), the normalized CPU costs of Redo Logging and Read After Write are on average $1.89$ and $1.96$ times. However, for update-only workload (100\% write), the benefits of using Erda are relatively small compared to those obtained with other three workloads. The normalized CPU costs of Redo Logging and Read After Write with update-only workload are on average $1.17$ and $1.11$ times.

\subsection{Log Cleaning}

\begin{figure}[t]
  \centering
    \includegraphics [width=0.48\textwidth]{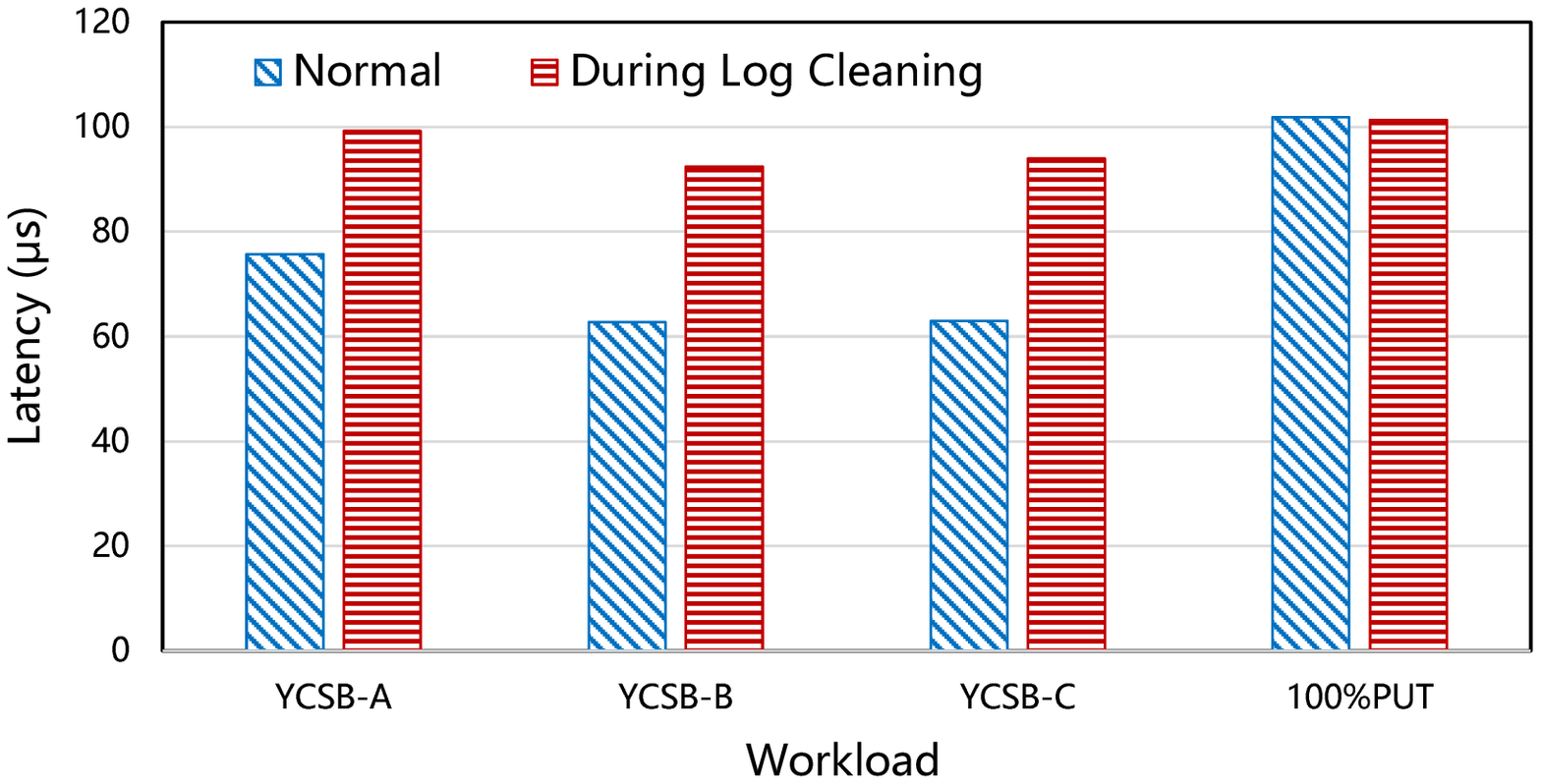}
\vspace{-7px}
    \caption{\label{fig:LogCleaningLatency1024B} The average latencies with four YCSB workloads when the value size of the key-value pair is $1,024$Bytes.}
  \vspace{-7px}
\end{figure}

As described in Section~\ref{subsec:Log Cleaning}, a server performs log cleaning and handles read/write requests concurrently. We evaluate the impact of log cleaning on the concurrent read/write requests. Figure~\ref{fig:LogCleaningLatency1024B} shows the average latencies of read/write requests under the normal cases of Erda and that of read/write requests during log cleaning, respectively.  We use four YCSB workloads, and the value size of the key-value pair is $1,024$Bytes. From Figure~\ref{fig:LogCleaningLatency1024B}, we observe that the highest average latency of read/write requests during log cleaning comes from using update-only workload. However, for update-only workload, the average latency during log cleaning is approximate to that of read/write requests under the normal cases of Erda. For YCSB-C workload (100\% read), the average latency of read/write requests during log cleaning is worse than that of read/write requests under the normal cases of Erda. The main reason is that the read procedure of Erda does not involve server CPUs with one-sided RDMA read, while the read procedure during log cleaning uses two-sided RDMA send (similar to Redo Logging and Read After Write).

\begin{table}[!hbp]
\renewcommand\arraystretch{1.5}
\vspace{-6px}
\caption{\label{table1} The Number of NVM writes in different operations. $N$ is the size of one key-value pair. Size(key) is the size of the key.}
\vspace{7px}
\centering
\arrayrulewidth0.8pt
\small
\begin{tabular}{|c|c|c|c|}
\hline \tabincell{c}{NVM Writes\\ (Bytes)}&Create&Update&Delete\\
\hline
\hline Erda&\tabincell{c}{Size(key)+\\10+N}&9+N&Size(key)+9\\
\hline Redo Logging&\tabincell{c}{Size(key)+\\12+2N}&4+2N&Size(key)+8\\
\hline Read After Write&\tabincell{c}{Size(key)+\\12+2N}&4+2N&Size(key)+8\\
\hline
\end{tabular}
\vspace{-2px}
\end{table}

\subsection{The Number of NVM Writes}

Table~\ref{table1} shows the number of NVM writes in create, update and delete operations. $N$ is the size of one key-value pair. Size(key) is the size of the key. In Erda, one create operation needs to first write metadata in an entry of a hash table in a server. Specifically, the server writes an object key, a head ID ($1$Byte), a new tag and an offset ($4$Bytes) that belong to an $8$-byte atomic write region in metadata. Then a client directly writes an object ($5$Bytes+$N$) in a log region. Therefore, the number of NVM writes is Size(key)+$10$Bytes+$N$. For an update operation in Erda, the server rewrites a new tag and an offset ($4$Bytes) in metadata, and then a client writes the updated object ($5$Bytes+$N$) in a log region. Therefore, the number of NVM writes is $9$Bytes+$N$. A delete operation in Erda is similar to an update, except that a delete object to be written in a log region is $5$Bytes+Size(key). Therefore, the number of NVM writes is Size(key)+$9$Bytes.

The number of NVM Writes in Redo Logging and Read After Write are the same. For a create operation, a server writes the metadata in a hash table with a key and an address ($8$Bytes). Then a key-value pair and a CRC checksum ($4$Bytes) are written in the ring buffers (Read After Write) or redo log regions (Redo Logging). At last, the server verifies the integrity of the key-value pair, and then writes the key-value pair to the destination address. Therefore, the number of NVM writes is Size(key)+$12$Bytes+$2N$. For an update operation, a server does not update the metadata in a hash table. The procedure of writing a key-value pair to the destination address follows the operations of the create. Therefore, the number of NVM writes is $4$Bytes+$2N$. For a delete operation, a server sets the metadata in a hash table to $0$, but does not write data on the destination address of a key-value pair. Therefore, the number of NVM writes is Size(key)+$8$Bytes.

In summary, compared with Redo Logging and Read After Write, Erda reduces NVM writes approximately by $50\%$, while significantly decreasing latency, improving throughput and reducing CPU consumption.

\section{Related Work}\label{sec:related work}

\textbf{Consistency guarantee for RDMA-based NVM.} Currently, providing persistence and consistency guarantees for RDMA writes to NVM typically requires extra network round-trips or CPU participation~\cite{yang2019orion,ChetDouglas_RDMA_with_PM}.  For example, a general method for providing these guarantees is to follow RDMA write(s) with an RDMA read to force client data to Asynchronous DRAM Refresh (ADR) domain, or to follow RDMA write(s) with an RDMA send to obtain local callback and persistency~\cite{ChetDouglas_RDMA_with_PM,ChetDouglas_Intel_Perspective}. HyperLoop~\cite{kim2018hyperloop} offloads replicated transactions to RDMA NICs by programming RNICs in multi-tenant storage systems, with NVM as a storage medium. This paper designs a new RDMA FLUSH (gFLUSH) primitive to support the durability at the NIC-level. However, gFLUSH essentially leverages an extra RDMA read operation, thus increasing network round-trips. Moreover, HyperLoop doesn't consider the NVM lifetime (the write operations are not optimized). If a failure occurs during the transaction, this transaction will be abandoned without recovery. Orion~\cite{yang2019orion}, a distributed file system for NVMM-based storage, ensures persistence by CPU involvement, thus providing remote data atomicity. DSPM~\cite{shan2017distributed} proposes a kernel-level distributed persistent memory system that integrates distributed memory caching and data replication techniques. DSPM guarantees crash consistency both within a single node and across distributed nodes with CPU involvement. Mojim~\cite{zhang2015mojim} uses a primary-backup protocol to replicate PM data through two-sided RDMA. It provides consistency and durability guarantees with CPU participation. Unlike existing schemes, we provide persistence and consistency guarantees for one-sided RDMA writes to NVM without extra network round-trips or remote CPU consumption. Moreover, compared with existing consistency mechanisms such as undo/redo logging and copy-on-write, we also reduce the NVM writes.

\textbf{System Optimizations for RDMA-based NVM.} Octopus~\cite{lu2017octopus} is RDMA-enabled persistent memory system, which proposes the local logging with remote in-place update for crash consistency. However, the solution fails to ensure remote data atomicity. Specifically, in Collect-Dispatch transaction of Octopus, a coordinator uses one-sided RDMA write to update the write sets in participants, and thus participants are unaware of the incomplete data without the CPU involvements of participants. Persistence Parallelism Optimization~\cite{hu2018persistence} improves the parallelism of maintaining the orders for write requests in the memory bus and the RDMA network. NVFS~\cite{islam2016high} is an optimized HDFS with NVM and RDMA. It re-designs HDFS I/O with memory semantics to exploit the byte-addressability of NVM. ScaleRPC~\cite{chen2019scalable} is an efficient RPC primitive to alleviate resource contention and achieve high scalability. It introduces connection grouping and virtualizes the mapping with one-sided RDMA verbs on RC (reliable connection). However, these RDMA-based NVM systems do not provide solutions for remote data atomicity. Unlike them, our proposed Erda is to guarantee remote data atomicity without extra network round-trips, remote CPU consumption and redundant copy.

\section{Conclusion}\label{sec:conclusion}

In order to address the problems of high network overheads, high CPU consumption and double NVM writes when ensuring remote data atomicity under RDMA and NVM scenarios, we propose a zero-copy log-structured memory design, called Erda. Erda guarantees remote data atomicity without extra network round-trips, remote CPU consumption and redundant copy. It transfers data directly to the destination address without buffer and copy, and guarantees consistency and atomicity by leveraging out-of-place updates, CRC checksum and $8$-byte atomic write. Evaluation results demonstrate that Erda reduces NVM writes approximately by $50\%$, as well as significantly reduces CPU cost, decreases latency and improves throughput. We have released the source codes for public use at \emph{https://github.com/csXinxinLiu/Erda}.

\end{document}